\documentclass[10pt,preprint,a4paper]{aastex}
\usepackage{graphics,epsf}
\usepackage{amsmath}                % American Mathematical Society package
\usepackage{amsfonts}               % American Mathematical Society fonts
\usepackage{amssymb}                % American Mathematical Society symbol
\usepackage{epsfig}                 % EPS figures
\usepackage{graphicx}
\usepackage{float}

\newcommand{\cm}{{~\rm cm}}
\newcommand{\km}{{~\rm km}}
\newcommand{\s}{{~\rm s}}

\newcommand{\erg}{{~\rm erg}}

\begin{document}

\title{The two promising scenarios to explode core collapse supernovae}

\author{Noam Soker\altaffilmark{1}}

\altaffiltext{1}{Department of Physics, Technion -- Israel Institute of Technology, Haifa
32000, Israel; soker@physics.technion.ac.il}

%\label{firstpage}
%\maketitle

\begin{abstract}
I compare to each other what I consider to be the two most promising scenarios to explode core-collapse supernovae (CCSNe). Both are based on the negative jet feedback mechanism (JFM). In the jittering jets scenario a collapsing core of a single slowly-rotating star can launch jets. The accretion disk or belt (a sub-Keplerian accretion flow concentrated toward the equatorial plane) that launches the jets is intermittent with varying directions of the axis. Instabilities, such as the standing accretion shock instability (SASI), lead to stochastic angular momentum variations that allow the formation of the intermittent accretion disk/belt. According to this scenario no failed CCSNe exist. According to the fixed axis scenario, the core of the progenitor star must be spun up during its late evolutionary phases, and hence all CCSNe are descendants of strongly interacting binary systems, most likely through a common envelope evolution (whether the companion survives or not). Due to the strong binary interaction, the axis of the accretion disk that is formed around the newly born neutron star has a more or less fixed direction. According to the fixed axis scenario, accretion disk/belt are not formed around the newly born neutron star of single stars; they rather end in failed CCSNe.
{{{ I also raise the possibility that the jittering jets scenario operates for progenitors with initial mass of $ 8 M_\odot \la M_{\rm ZAMS} \la 18 M_\odot$, while the fixed axis scenario operates for $M_{\rm ZAMS} \ga 18 M_\odot$.  }}} For the first time these two scenarios are compared to each other, as well as to some aspects of neutrino-driven explosion mechanism. These new comparisons further suggest that the JFM plays a major role in exploding massive stars.
\end{abstract}

%%\begin{keywords}
%% Keywords: supernovae: general --- binaries: close --- stars: jets
%%\end{keywords}

% ==========================================================
\section{INTRODUCTION}
\label{sec:intro}
% ==========================================================

%% \bibitem[Langer(2012)]{2012ARA\&A..50..107L} Langer, N.\ 2012, \araa, 50, 107

Although we know what is the most popular explosion mechanism of core collapse supernovae (CCSNe) in the scientific literature, we are yet to find the most popular mechanism among exploding stars in nature.
The delayed neutrino mechanism (e.g., \citealt{Bruennetal2016, Jankaetal2016, Muller2016, Burrowsetal2017}), although the most popular in the literature, encounters problems (e.g., \citealt{Papishetal2015, Kushnir2015b}). An alternative mechanism to account for all CCSNe is the jet feedback mechanism (JFM; for a review see \citealt{Soker2016Rev}).

Observations and their analysis of supernova remnants (SNRs) and of polarizations in CCSNe (e.g., \citealt{Wangetal2001, Maundetal2007, Lopezetal2011, Lopezetal2013, Lopezetal2014, Milisavljevic2013, Gonzalezetal2014, FesenMilisavljevic2016, Inserraetal2016, Mauerhanetal2017, GrichenerSoker2017, BearSoker2017, Bearetal2017}) suggest that jets play a role in at least some CCSNe. As well, several arguments based on analytical studies and numerical simulations suggest that the collapse of a pre-explosion rapidly rotating core leads to the formation of two opposite well collimated jets (e.g. \citealt{Khokhlovetal1999, MacFadyen2001, Hoflich2001, Woosley2005, Burrows2007, Couch2009, Couch2011, TakiwakiKotake2011, Lazzati2012, Maedaetal2012, Mostaetal2014, Nishimura2015, BrombergTchekhovskoy2016, Gilkis2017, Nishimuraetal2017}).
The condition of a rapidly rotating core requires that a stellar binary companion enters the envelope and spirals-in to the core. Therefore, not all massive stars are expected to possess a rapidly rotating core when their core collapses. Indeed, most of these papers take it that jets are involved in only a small fraction of all CCSNe.

Although the notion that jets play a role in some CCSNe is old, the idea that all CCSNe are exploded by jets and that the jets operate via a negative JFM is relatively new (e.g., \citealt{PapishSoker2011}; review by \citealt{Soker2016Rev}). The main problem for scenarios that are based on the JFM is to supply the required angular momentum to form an accretion disk or an accretion belt around the newly born neutron star (or a black hole). The disk or the belt are required to launch the jets that explode the star.

Clearly a strongly interacting binary companion can deliver angular momentum to the envelope of the progenitor, and from there possibly to the core.  In addition, instabilities might lead to a stochastic accretion of gas with varying specific angular momentum, to the point that an intermittent accretion disk or a belt form.
In that case the scenario is termed the jittering jets scenario. The second possibility is that instabilities do not lead to the formation of an accretion disk or belt, and the star cannot explode if it is not spun-up by a companion. This is termed the fixed axis scenario.
These two contesting scenarios are distinguished by the still open question of whether instabilities can lead to a stochastic accretion process that forms intermittent accretion disks or belts that launch jets.

Although the JFM for exploding CCSNe was discussed in recent years, here for the first time I present the two scenarios as contesting scenarios (sections \ref{sec:jittering} and \ref{sec:fixed}) in the frame of the JFM. Also, for the first time the fixed axis scenario is presented under the JFM.

This study is motivated by new observations and their analysis (sections \ref{sec:observational} and \ref{sec:others}), and in part by claims against the jittering jets scenario.
I do note that the arguments against the jittering jets scenario are not strong, as indicated for example by their presentations only in footnotes, rather than in a long physical discussion (e.g.,  \citealt{Jankaetal2016}), or by using simulations that do not include magnetic fields (e.g., \citealt{Muller2016}).
Nonetheless, I do take these into account, and present for the first time the alternative fixed axis scenario in the frame of the JFM (section \ref{sec:fixed}), that accounts for the possibility that the neutron star that is formed by a single star progenitor cannot launch jets.
For the first time these two scenarios are compared to each other.
I also discuss these two scenarios in relation to other explosion mechanisms (section \ref{sec:others}) to better emphasize their advantages. I present there the first critical analysis of some recent claims for success of the delayed neutrino mechanism.
Along the paper and in the summary (section \ref{sec:summary}), I point out some simulations and observations that might break the tie between the two scenarios.

% ==========================================================
\section{THE JITTERING JETS SCENARIO}
\label{sec:jittering}
% ==========================================================

A crucial ingredient in the jittering jets scenario is that instabilities, before and after core collapse, can form an accretion disk or an accretion belt around the newly born neutron star or black hole.
An accretion belt is defined here to be a thick sub-Keplerian rotating accretion inflow that does not extend much beyond the neutron star (or black hole), and has sufficiently large specific angular momentum to prevent a dense inflow along the two opposite polar directions. \cite{SchreierSoker2016} suggest that such an accretion belt might launch jets.
In a series of paper Gilkis \& Soker \citep{GilkisSoker2014, GilkisSoker2015, GilkisSoker2016}
argue that the pre-collapse turbulence regions that exist in the core, and more so for the turbulence that was assumed and used by \cite{CouchOtt2013}, \cite{CouchOtt2015}, and \cite{MuellerJanka2015}, might lead to the formation of an intermittent accretion belt around the neutron star that is formed at the center of the collapsing core.

In addition to the pre-collapse turbulence, there are post-collapse instabilities in the post-shock inflow toward the neutron star, such as due to heating by neutrinos, and in particular the standing accretion shock instability (SASI; e.g., \citealt{BlondinMezzacappa2003, BlondinMezzacappa2007, Fernandez2010, Burrows1995, Janka1996, Buras2006a, Buras2006b, Ott2008, Marek2009, Iwakamietal2014, Abdikamalovetal2015, Fernandez2015}). Most relevant to the jittering jets scenario are the spiral modes of the SASI that carry local angular momentum variations. \cite{Rantsiouetal2011} and \cite{Kazeronietal2017}, for example, study the influence of the SASI modes on the final angular momentum of neutron star.

\cite{Papishetal2015} argue that the SASI can lead to the formation of an accretion belt. They use the results of \cite{Fernandez2010} for the variation with time of the specific angular momentum $j(t)$ of the accreted gas due to the SASI, and calculate the angle from the polar axis $\theta_a(t)$ within which mass possessing this specific angular momentum $j(t)$ cannot be accreted. They derive the expression
\begin{equation}
\begin{split}
\theta_a =
\sin^{-1} \sqrt{\frac{j_z(t)}{j_\mathrm{Kep}}} \simeq &
0.3
\left( \frac{j_z(t)}{2 \times 10^{15} \cm^2 \s^{-1}} \right)^{1/2} \\
& \times
\left( \frac{M_\mathrm{NS}}{1.4 M_\odot} \right)^{-1/4}
\left( \frac{R_\mathrm{NS}}{20 \km} \right)^{-1/4},
\end{split}
\label{eq:angle}
\end{equation}
where $j_\mathrm{Kep} = \sqrt{GM_\mathrm{NS}R_\mathrm{NS}}$, and $M_\mathrm{NS}$ and $R_\mathrm{NS}$ are the mass and radius of the newly born neutron star, respectively. In the second equality I approximated for a small angle $\theta_a$.

In light of the new calculations of the SASI by \cite{Kazeronietal2017}, I repeat the calculations
of \cite{Papishetal2015}. \cite{Kazeronietal2017} assume that the pre-collapse core is slowly rotating. Their calculations are not full three-dimensional ones, but rather done in cylindrical geometry. For that, their results are not final yet and full three-dimensional simulations are required to find the exact properties of SASI in pre-collapse slowly rotating cores.
The fluctuations of $j(t)$ are centered around the initial value of the pre-collapsing core $j_0$. In their calculations with an initial specific angular momentum of $j_0=10^{15} \cm^2 \s^{-1}$ and a ratio between the initial shock radius and the proto neutron star radius of R=3, the maximum value of $j(t)$ is about $3 \times 10^{15}\cm^2 \s^{-1}$ (R. Kazeroni, private communication). This value is the reason for the scaling used of $j(t)$ in equation (\ref{eq:angle}).
In an accretion time of one second, at about half of the time and in about ten episodes the specific angular has a value of $\vert j(t) \vert > 10^{15} \cm^2 \s^{-1}$. According to the jittering jets scenario, the result will be about ten jet-launching episodes that have enough energy to explode the star with an energy of about $10^{51} \erg$ \citep{PapishSoker2014a, PapishSoker2014b}.

Equation (\ref{eq:angle}) is derived under the assumption that the accreted gas has a uniform specific angular momentum. But the value of $j(t)$ is not uniform for the accreted gas.
Gas with lower angular momentum than $j(t)$ might flow through the poles with an angle $\theta<\theta_a$, while gas with higher specific angular momentum will form a flatter accretion belt, i.e., will have a value of $j_{\rm belt}(t) > j(t)$. The limiting angle $\theta_a$ in equation (\ref{eq:angle}) represents some typical behavior.

Two additional processes act to increase the value of the opening angle along the polar directions. These are ($i$) magnetic field amplification in the accretion belt, i.e., a dynamo, and ($ii$) neutrino heating.

\cite{SchreierSoker2016} crudely estimate the amplification of the magnetic field in an accretion belt in a non-turbulent region. Their estimate can be written for the magnetic pressure in the disk as
\begin{equation}
\begin{split}
\frac {P_B}{\rho v^2_{\rm esc}}
\approx  \left( \frac{j_{\rm belt}}{j_\mathrm{Kep}} \right)^2 \approx 0.01  &
\left( \frac{j_{\rm belt}}{2 \times 10^{15} \cm^2 \s^{-1}} \right)^{2} \\
& \times
\left( \frac{M_\mathrm{NS}}{1.4 M_\odot} \right)^{-1}
\left( \frac{R_\mathrm{NS}}{20 \km} \right)^{-1}.
\end{split}
\label{eq:Pb}
\end{equation}
This is a non-negligible ratio when the following points are considered. (1) This estimate is for the non-turbulent regions of an accretion belt. As turbulence is expected, the amplification will be much more efficient. Overall, the magnetic field will be stronger than the estimated value in equation (\ref{eq:Pb}).
(2) Out of the accreted $\approx 0.1 M_\odot$ in the final mass accretion period, it is sufficient that about five to ten per cents of that mass be ejected at the escape speed to supply an explosion energy of about $10^{51} \erg$. Also, magnetic field activity, such as reconnection, is likely to expel some mass from the polar directions, hence increasing the opening angle along the polar directions \citep{SchreierSoker2016}.
(3) \cite{Endeveetal2012} find that the SASI can substantially increase the strength of the magnetic fields outside the neutrinosphere. This implies that the initial magnetic field of the matter accreted in the disk is large.

Overall, the value of the magnetic energy in the accretion belt can be tens of per cents of the binding energy of the gas, which is sufficient to eject about 5-10 percents of the mass in the belt and explode the star.

The second effect that can increase the opening angle along the polar directions is heating by neutrinos. Simulations of core collapse with rotation show that the lower density inflow along the polar directions has higher entropy (e.g., \citealt{Gilkis2017}).
\cite{Kotakeetal2003} find that neutrino heating is stronger near the rotational axis than near the equatorial plane, and suggest that this might expel mass more efficiently along the polar directions. Their calculations are for matter outside the neutrinosphere, while the accretion belt studied here is within the neutrinosphere. Nonetheless, the ejection of mass along the polar directions outside the neutrinosphere will reduce the mass inside the neutrinosphere along the same directions.

Overall, my conclusion is that fluctuations, such as core turbulence before collapse and the SASI after collapse, aided by neutrino heating along the angular momentum axis and amplification of magnetic fields by the SASI and in the accretion belt, are likely to form an accretion belt that is likely to launch jets even in pre-collapse very slowly rotating cores.

% ==========================================================
\section{THE FIXED AXIS SCENARIO}
\label{sec:fixed}
% ==========================================================

If, despite the conclusion at the end of section \ref{sec:jittering}, stars that are not spun-up along their evolution cannot launch jets, i.e., the jittering jets scenario does not work, then the JFM mechanism requires that the progenitors of all CCSNe are spun-up by a binary companion. The collapsing core will have a large amount of angular momentum, and the angular momentum axis will be, more or less, fixed. The instabilities discussed in section \ref{sec:jittering} will cause small variations in the direction. The mass loss rate into the jets might change, but not the direction of the jets.

In the fixed axis scenario the amount of angular momentum deposited by a binary companion, $J_{\rm dep}$, should be at least as large as the maximum angular momentum that the single star can possess on the main sequence, $J_{\rm MS, max}$.

A star that along its entire evolution does not acquire angular momentum from a stellar companion or a sub-stellar companion, or that the angular momentum it acquires is less than the maximum value it can have on the main sequence, is termed an angular momentum isolated star, or
an J-isolated star \citep{SabachSoker2017}
\begin{equation}
J_{\rm dep}\la J_{\rm MS, max} \qquad {\rm for~ a~ J-isolated~ star} .
\label{eq:jsoalted}
\end{equation}
J-isolated stars do not correspond one-to-one with single stars. Binary stars with a large orbital separation, such that the companion does not spin-up the primary star, are J-isolated stars. Single low mass stars that have close and massive planets can become non-J-isolated stars if they are spun-up by such a planet to the degree that $J_{\rm dep} \ga J_{\rm MS, max}$ \citep{SabachSoker2017}.

If the jittering jets scenario does not work, therefore, all progenitors of CCSNe are non-J-isolated stars. As the progenitors are massive stars, they must strongly interact with a binary companion. It is not clear whether a binary companion outside the envelope of a giant star can spin-up the core of the progenitor to the required degree. It might be, then, that in the fixed axis scenario all progenitors of CCSNe are not only non-J-isolated stars, but are all a product of a common envelope evolution, e.g., as the accepted model for the progenitor of SN~1987A and as was recently claimed for the progenitor of the SNR RCW 86 \citep{Gvaramadzeetal2017}.
{{{ For that matter, the companion can survive or not the common envelope evolution. Most of the angular momentum is deposited to the envelope as the companion spirals-in from the surface to a very small radius. The disk that launches the jets during the explosion is formed from the collapsing iron or silicon core, and the demand is that the core rotates rapidly. The new accretion disk around the neutrino star (or black hole) is not related to a possible accretion disk that might be formed around the core as a result of the destruction of the companion at the end of the common envelope evolution.  }}}

% ==========================================================
\section{OBSERVATIONAL CONSEQUENCES}
\label{sec:observational}
% ==========================================================

According to the JFM in many supernovae there should be signatures of jets, as indeed observed (see section \ref{sec:intro}).
However, in all of these cases both scenarios can account for the properties of the jets. One example is the presence of `ears' in SNRs (e.g., \citealt{Bearetal2017}). In most of these cases the required energy to inflate the ears is only $\approx 5-15 \%$ of the kinetic energy of the SNR \citep{GrichenerSoker2017, Bearetal2017}. According to the JFM only the jets that are blown at the end of the process can leave clear signatures on the morphology of the SNR. At early times of the explosion process the jets are stopped in the core and the inner regions of the star, and by that explode the star. This range of energies for the inflation of the  ears is expected in one jet-launching episode of the jittering jets scenario. In cases where the energy is much larger, e.g., as in the SNR W49B \citep{BearSoker2017}, an explosion with a fixed axis occurs. But according to the jittering jets scenario some CCSNe do occur after their progenitor was spun-up by a companion in a common envelope evolution, and a fixed axis is expected. So at present, signatures of jets cannot break the tie between the two scenarios.

\cite{Tanakaetal2017} suggest from their modelling of line polarization of CCSNe that ``. . .  SN ejecta may have an overall 2D bipolar structure inside and 3D clumpy structure outside.'' As I discuss later, one of the expectation of the JFM is the presence of global axisymmetrical bipolar structure together with instabilities.
However, observations of polarization that indicate asymmetrical explosion cannot distinguish between the two scenarios discussed here. In both scenarios it is expected than many CCSNe will be axisymmetrical (bipolar), and in both scenarios it is expected that many CCSNe will have bipolar circumstellar matter (CSM). In the fixed axis scenario all progenitors have gone through a strong binary interaction, and in the jittering jets scenario a large fraction of them did so.

According to the jittering jets scenario in some explosions the two opposite jets from the last two jet-launching episodes might leave signature in the SNR. In that case the SNR possesses two pairs of ears along different axes. However, at this time even such a signature cannot break the tie between the two scenarios. In the fixed axis scenario, two pairs of ears in the SNR along different axes might result from the presence of ears in the CSM before the explosion. Such a mechanism for the formation of ears was proposed for ears in Type Ia SNe that exploded inside a CSM with ears (e.g., \citealt{TsebrenkoSoker2015}; they are termed SNIP, for SN inside planetary nebulae).

According to the jittering jets scenario there are no failed CCSNe. Even if the inner layers of the core do not explode the star, the convective helium layer has large fluctuations of angular momentum before collapse, and stochastic accretion will lead to explosion \citep{GilkisSoker2016}. In the fixed axis scenario J-isolated stars do not explode, and hence result in failed SNe. At present there is no clear case for a failed SN, but rather different suggestions, e.g., of fast radio bursts \citep{Katz2017}.
\cite{Adamsetal2017} suggest that the star N6946-BH1 that erupted in 2009 \citep{Gerkeetal2015} is a failed SN, as its behavior is similar to a failed SN model \citep{Nadezhin1980, LovegroveWoosley2013}. However, \cite{KashiSoker2017} propose that this event is a Type II intermediate luminosity optical transient (ILOT). In a type II ILOT the strongly interacting binary system that powers the ILOT ejects mass in the equatorial plane that blocks the central source from our line of sight. So this specific event cannot yet rule out the jittering jets scenario.

It seems that the pre-explosion outbursts of CCSNe are common (e.g., \citealt{Moriyaetal2014, Ofeketal2014, SvirskiNakar2014, Tartagliaetal2016, Yaronetal2017}), with about one in ten CCSNe suffering a pre-explosion outburst (e.g., \citealt{Marguttietal2017}).
The outburst might result from a single-star process, e.g., strong convection in the pre-collapsing core \citep{QuataertShiode2012, ShiodeQuataert2014}. It might as well result from both instability in a single star, but enhanced by a binary interaction (e.g., \citealt{McleySoker2014}).
One possibility is that the outburst starts with a dynamo activity in the core \citep{SokerGilkis2017}. The dynamo requires both strong convection and rapid rotation.
The occurrence of pre-explosion outbursts cannot yet be used to distinguish between the two scenarios studied here because we cannot determine the instability mechanism inside the core. But future three-dimensional magneto-hydrodynamic simulations (that are highly demanding) might shed light on the required pre-collapse core rotation. In any case, I argue that strong pre-explosion outbursts require the presence of a binary companion.

% ==========================================================
\section{RELATION TO OTHER EXPLOSION MECHANISMS}
\label{sec:others}
% ==========================================================
% ======================
\subsection{Ejecta distribution}
\label{subsec:distribution}
% ======================

\begin{figure} % [b]
% \vspace*{-2.0 cm}
\begin{center}
\vskip -0.20 cm
 \includegraphics[width=3.8in]{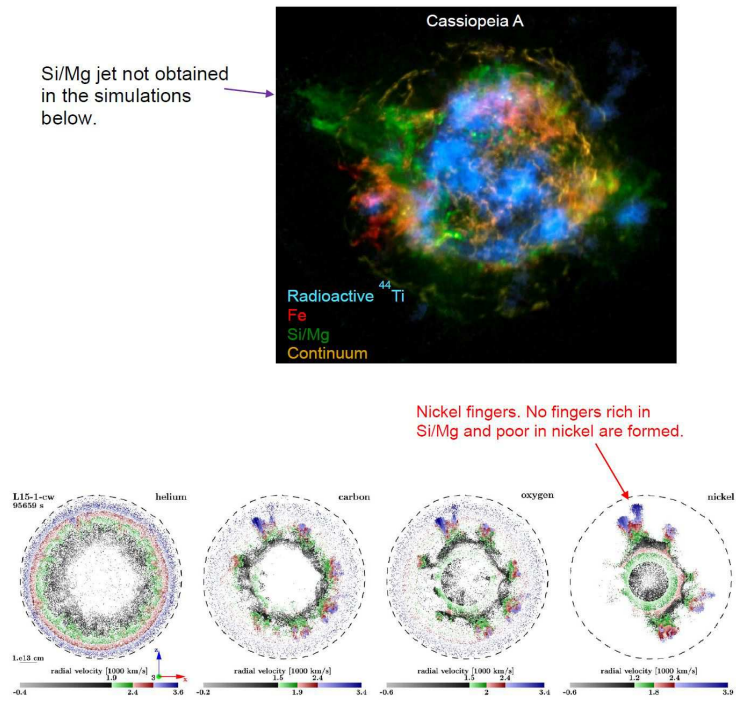}
\vskip 0.50 cm
 \caption{Upper panel: The observation of Cassiopeia A (from \citealt{Grefenstetteetal2017}). Lower panel:  Numerical simulations based on the neutrino-driven explosion mechanism (from \citealt{Wongwathanaratetal2015}). It is clear that the simulations do not reproduce all properties of the SNR, such as the jet.  }
 \label{fig:CassiopeiaA}
\end{center}
\end{figure}

\cite{Grefenstetteetal2017} present the distribution of $^{44}$Ti in the Cassiopeia A SNR (also \citealt{Leeetal2017}). \cite{Wongwathanaratetal2015} present numerical simulations based on neutrino-driven explosion, and argue that they reproduce the protrusions and the distribution of some metals in Cassiopeia A. In Fig. \ref{fig:CassiopeiaA} I present these observations and numerical  simulations.
Based on a critical comparison of the observations and simulations, I argue that the numerical results do not explain the observations of the north-east jet of Cassiopeia A. ($i$) The (north-east) jet in Cassiopeia A is Si-rich, and does not seem to contain iron (which is the product of nickel). ($ii$) The instability-fingers formed by the nickel in the numerical simulations do not move outward much faster than the (helium-rich) main shell of the supernova. Such fingers cannot form protrusions extending outside the main shell of the SNR. There are instabilities, as also expected in the JFM, but they are not sufficient to explain the jets in Cassiopeia A. Indeed, \cite{Orlandoetal2016} had to introduce large-scale anisotropies (that I attribute to jets) to reproduce the structure of Cassiopeia A.
%FFFFFFFFFFFFFFFFFFFFFFFFFFFFFFFFFFFFFFFFFFFFFFFFFFFFFFFFFFFFFFFFF
%FFFFFFFFFFFFFFFFFFFFFFFFFFFFFFFFFFFFFFFFFFFFFFFFFFFFFFFFFFFFFFFFF

\cite{Grefenstetteetal2014} propose that the $^{44}$Ti nonuniform distribution in Cassiopeia A result from a multimodal explosion, such as expected from instabilities. But as noted above instabilities do not reproduce the jets of Cassiopeia A.
In \cite{Gilkisetal2016} we discuss the jittering jets scenario and the $^{44}$Ti distribution in Cassiopeia A. We argue there that the jittering jets scenario has the properties of a multimodal explosion, because several pairs of opposite jets are launched in different directions.

In the left panel of Fig. \ref{fig:SN1987A} I present the iron distribution in SN 1987A taken from \cite{Larssonetal2016}, alongside numerical simulations taken from \cite{Wongwathanaratetal2015}.
While in Cassiopeia A instabilities alone cannot explain the Si-rich jet, in SN 1987A instabilities might in principle account for the iron structure. However, the match between observations and simulations based on the neutrino-driven explosion is not satisfactory.
The simulations lead to narrow Ni-rich (later turn to Fe) fingers, while the observed iron distribution in SN 1987A is concentrated in two approximately opposite wide regions. I argue that large asymmetries, such as jets, must be introduced in addition to the instabilities. When new ALMA observations of SN 1987A \citep{Matsuuraetal2015} are considered, the need for jets becomes clearer even.
It is important to note that the concentration of iron near the equatorial plane of the ring does not contradict the JFM . In \cite{BearSoker2017} we present arguments for the shaping of the CCSN remnant W49B by jets that were launched perpendicular to the iron-rich stripe in that SNR (for more details see also {\citealt{Soker2017Proc}).
% FFFFFFFFFFFFFFFFFFFFFFFFFFFFFFFFFFFFFFFFFFFFFFFF
\begin{figure}% [b]
\begin{center}
\includegraphics[scale=0.62]{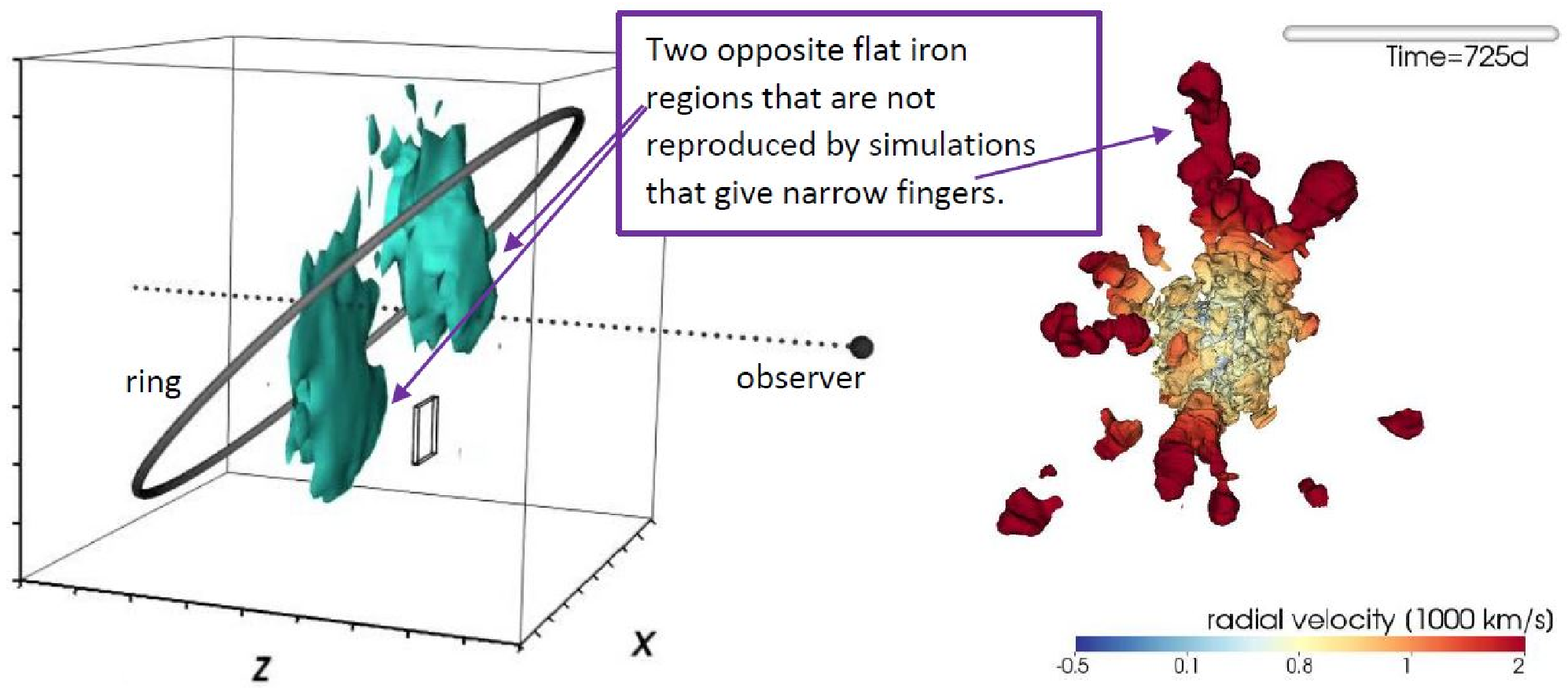}
\vskip +0.5 cm
\caption{Left panel: Observed Fe-morphology in SN 1987A (from \citealt{Larssonetal2016}). Right panel: Results of numerical simulations based on neutrino-driven explosion (from \cite{Wongwathanaratetal2015}). The numerical simulations form narrow fingers that do not cover all properties of the Fe-rich regions.  }
\label{fig:SN1987A}
\end{center}
\end{figure}
% FFFFFFFFFFFFFFFFFFFFFFFFFFFFFFFFFFFFFFFFFFFFFFFF

I can summarize this subsection as following.
The JFM, that includes both jittering jets and instabilities, seems to better explain the ejecta distributions in the SNRs Cassiopeia A and SN 1987A than the neutrino-driven mechanism does.

% ======================
\subsection{Energy}
\label{subsec:energy}
% ======================

As mentioned in section \ref{sec:intro}, the most popular explosion mechanism in the literature is the delayed neutrino mechanism. But even its supporters admit that this mechanism cannot yield CCSN explosion (kinetic) energies of $E_{\rm exp} \ga 2 \times 10^{51} \erg$ (e.g., \citealt{Fryer2006, Fryeretal2012, Sukhboldetal2016, SukhboldWoosley2016}), and hence cannot account for super-energetic CCSNe. Super-energetic CCSNe can reach energies of well above $10^{52} \erg$, and their study is a hot topic (e.g., \citealt{GalYam2012, Moriyaetal2015, Wangetal2016, Arcavietal2016, Sorokinaetal2016, Abbottetal2017}). In the delayed neutrino mechanism there is a need for an additional energy source. Most popular in the literature is a strongly magnetized rapidly-rotating neutron star (a magnetar, e.g., \citealt{Metzgeretal2015}).
The formation of a magnetar would most likely be accompanied by jets that carry much more energy than $2 \times 10^{51} \erg$, and possibly more than is stored in the newly born magnetar \citep{Soker2016Mag}. So the formation of a magnetar makes the delayed neutrino mechanism a negligible process in that case. If late accretion takes place, the jets can carry even more energy and for a longer time (e.g., \citealt{Gilkisetal2016}).

Overall, the JFM is compatible with super-energetic CCSNe, and nicely account for them \citep{Soker2017}.
In any case, the formation of a magnetar, if occurs, requires a rapidly rotating pre-collapse core.
But such rare cases are expected in both JFM scenarios studied in the present paper, and hence  magnetars, or more generally super-energetic CCSNe, do not prefer one scenario over the other.

% ======================
\subsection{Angular momentum}
\label{subsec:AM}
% ======================

In the collapse-induced thermonuclear explosion (CITE) mechanism a mixed layer of helium and oxygen suffers a thermonuclear burning when it collapses and heats up, and by that it supposes to explode the star \citep{Burbidgeetal1957, KushnirKatz2014}. For that to occur the pre-collapse core must have a large amount of angular momentum \citep{Kushnir2015a}. As a result of that a very massive  accretion disk is formed around the newly born neutron star (\citealt{Gilkisetal2016}; confirmed later by \citealt{BlumKushnir2016}). The energy that is carried by the jets that this accretion disk launches is larger than the energy released by the thermonuclear reactions (\citealt{Gilkisetal2016}). The CITE mechanism requires all CCSN progenitors to be non-J-isolated stars. It actually requires a much larger amount of angular momentum of the pre-collapse core than the fixed axis scenario requires. Because of the angular momentum requirement that is stronger for the CITE mechanism than for the fixed axis scenario, and because of the expectation that jets will release more energy than the thermonuclear burning, I think that the CITE mechanism does not really help in exploding stars.

% ==========================================================
\section{DISCUSSION AND SUMMARY}
\label{sec:summary}
% ==========================================================

In this paper I compared for the first time the two scenarios that might operate within the JFM. Furthermore, it is the first time that the fixed-axis scenario is discussed in the frame of the JFM. In Table 1 I summarize the basic ingredients, demands, and consequences of the two scenarios that I discussed in previous sections.
It is important to note that in the jittering jets scenario many of the progenitors do go through a strong binary interaction, and in that sense behave like the CCSNe of the fixed axis scenario.
But the fixed axis scenario requires that many more massive stars experience a strong binary interaction, most likely a common envelope evolution.
% TTTTTTTTTTTTTTTTTTTTTTTTTTTTTTTTTTTTTTTTTTTTTTTTTTTTTTTTTTTTTTTTTTTTTTTT
%\begin{table}[H]
\begin{table*}
%\tiny
\label{Tab:Table1}
\begin{center}
\begin{tabular}{|c|c|c|}
  \hline
  Property & Jittering jets scenario & Fixed axis scenario \\
\hline
\hline
  Source of             & Binary interaction        & Binary interaction    \\
 angular momentum       & and/or instabilities      &     \\
\hline
 Axis of jets           & Might jitter              & Fixed in direction \\
\hline
 Demands                & (1) Violent instabilities & (1) Almost all massive stars  \\
                        & at collapse               & are non-J-isolated, mostly\\
                        & (2) Accretion belts can   & through common envelope \\
                        & launch jets               & interaction  \\
\hline
Black hole formation    & Inefficient JFM (because  & Inefficient JFM   \\
                        & of well collimated jets)  & or J-isolated stars \\
\hline
  Failed CCSNe          &  Do not exist             & From J-isolated stars  \\
\hline
Super energetic CCSNe   & Inefficient JFM           & Inefficient JFM   \\
and gamma ray bursts    & and late accretion        & and late accretion \\
\hline
 Implications           & All massive stars in      & (1) All CCSNe come from strongly     \\
                        & all masses explode        &  interacting binary systems   \\
                        &                           & (2) Bipolar CSM is common    \\
\hline
Supporting observations & Multiple ears             & (1) Bipolar CSM in some    \\
                        & in some SNRs              & SNRs (e.g., SN~1987A)   \\
                        &                           & (2) Many type Ib and Ic \\
                        &                           & CCSNe explode with energies \\
                        &                           & as of type II CCSNe \\
\hline
Required calculations   & 3D magneto hydrodynamics  & Population synthesis of \\
                        & simulations of CCSNe      & common envelope evolution \\
                        & with very high resolutions& of CCSN progenitors   \\
\hline
\end{tabular}
\end{center}
 \caption{Properties and implications of the two scenarios that are compared in this paper.
 JFM: jet feedback mechanism; CSM: circumstellar medium; J-isolated stars are defined in equation (\ref{eq:jsoalted}). }
\end{table*}
% TTTTTTTTTTTTTTTTTTTTTTTTTTTTTTTTTTTTTTTTTTTTTTTTTTTTTTTTTTTTTTTTTTTTTTTT

To support the fixed axis scenario it is important to show that a sufficient number of massive stars go through the common envelope evolution. As well, it is important to show that there are enough massive stars that despite the common envelope evolution retain most of their hydrogen-rich envelope, as some CCSNe explode as massive red giants. SN~1987A did go through a common envelope evolution and retained a large fraction of its hydrogen-rich envelope, but the progenitor became a blue star before explosion. Very detailed population synthesis studies are required, and a careful comparison with the distribution of different kinds of CCSN types (II, Ib, Ic, etc).

Many type Ib and Ic that have lost all their hydrogen envelope are thought to result from a common envelope evolution (e.g., \citealt{Yoon2015}). Many of them explode with energies similar to those of Type II CCSNe. This shows that the common envelope evolution can lead to regular CCSNe as far as energy is concerned. The same goes to SN~1987A that was a Type II SN of a blue giant, and had a typical explosion energy. This supports to some degree, or at least does not contradict with, the fixed axis scenario.

The crucial calculations to do are 3D magneto hydrodynamic simulations of the collapse process,
including a large volume of the pre-collapse core, and these simulations need to continue to the stage of the accretion of the helium layer (if an explosion does not take place first).
Such simulations are highly resource-demanding, and are on the limit of the possibility on the best computers. \cite{Mostaetal2015}, for example, performed simulations of CCSNe with pre-collapse rapidly rotating cores at very high resolutions. They found rapidly rotating material around the newly born neutron star, and that this material amplifies magnetic fields. But only in their simulations with very high spatial resolutions they obtained a large magnetic field amplification.
In their simulations the magnetic energy density becomes about equal to the turbulent energy density (equipartition).  Magneto hydrodynamic simulations of non-rotating cores with even higher resolutions are required to examine the feasibility of the jittering jets scenario.

{{{ One other, compromising, possibility should be considered. Observations might suggest that not all massive stars with a zero age main sequence mass of $M_{\rm ZAMS} \ga 18 M_\odot$ explode (e.g., \citealt{Smartt2015}).
If holds, then the JFM might account for that observation with the jittering jets mechanism operating for stars with $ 8 M_\odot\la M_{\rm ZAMS} \la 18 M_\odot$, and the fixed axis scenario operating for $M_{\rm ZAMS} \ga 18 M_\odot$. Namely, all the exploding stars with $M_{\rm ZAMS} \ga 18 M_\odot$ explode after a common envelope evolution or a grazing envelope evolution. In a recent paper I suggest that most of the progenitors of Type IIb supernovae experience the grazing envelope evolution before they explode \citep{Soker2017IIb}. Hence, these CCSNe, for example, might result also from stars with $M_{\rm ZAMS} \ga 18 M_\odot$  }}}

The present study can be summarized as follows. The explosion mechanism of CCSNe is not determined yet. For that, we should be open to all explosion mechanisms that were not ruled out yet, and compare them with each other. In the present study I argued that if the explosion mechanism is the negative JFM, then the evolutionary scenario should be one of the two that are summarized in Table 1. This is the first time these two scenarios are compared to each other. As well, this is the first time the JFM is compared favorably with the delayed neutrino mechanism in explaining the metal distribution in two SNRs.
 These comparisons add a small but significant support to the JFM.

I thank Avishai Gilkis for many helpful discussions and comments. I thank an anonymous referee for encouraging comments.
This research was supported by the Israel Science Foundation.

\label{lastpage}
\end{document}